\documentclass[conference]{IEEEtran}

\usepackage[utf8]{inputenc}
\usepackage{amsmath,amssymb,amsfonts}
\usepackage{booktabs}
\usepackage{microtype}
\usepackage{graphicx}
\usepackage{hyperref}
\usepackage{cite}

\usepackage{tikz}
\usetikzlibrary{arrows.meta, positioning, shapes.geometric}
\usepackage{pgfplots}
\pgfplotsset{compat=1.18}

\title{Entropy-Based Evidence for Bitcoin’s Discrete Time Mechanism}

\author{
    \IEEEauthorblockN{Bin Chen\IEEEauthorrefmark{1}\thanks{Corresponding author. Email: bchen@szu.edu.cn}}
    \IEEEauthorblockA{
        School of Electronics and Information Engineering,\\
        Shenzhen University\\
        bchen@szu.edu.cn
    }
    \and
    \IEEEauthorblockN{Pan Feng}
    \IEEEauthorblockA{
        School of Electronics and Information Engineering,\\
        Shenzhen University
    }
}

\begin{document}

\maketitle

\begin{abstract}
Bitcoin derives a verifiable temporal order from probabilistic block discovery
and cumulative proof-of-work rather than from a trusted global clock. 
We show
that block arrivals exhibit stable exponential behavior across difficulty
epochs, and that the proof-of-work process maintains a high-entropy search
state that collapses discretely upon the discovery of a valid block. This
entropy-based interpretation provides a mechanistic account of Bitcoin’s
non-continuous temporal structure. In a distributed network, however, entropy
collapse is not completed instantaneously across all participants. Using
empirical observations of temporary forks, we show that collapse completion
unfolds over a finite propagation-bounded interval, while remaining rapid in
practice.

\end{abstract}

\begin{IEEEkeywords}
Non-continuous time, entropy collapse, difficulty adjustment, proof-of-work.
\end{IEEEkeywords}

\section{Introduction}

A key challenge in decentralized systems is establishing a reliable notion of
time without centralized clocks. Chen~\cite{chen2025noncontinuous} proposed that
Bitcoin addresses this challenge by advancing time through stochastic block
discoveries rather than through an external timeline, and introduced the concept
of entropy collapse to describe the discrete resolution of uncertainty at block
discovery. While this framework clarifies the conceptual structure of Bitcoin’s
non-continuous time, it has not yet been supported by large-scale empirical
analysis. This paper fills that gap using real-world block arrival observations.

Existing studies of Bitcoin block timing largely focus on characterizing arrival
statistics or measuring network effects in isolation. Early work established
that proof-of-work produces approximately exponential inter-arrival times and
highlighted the impact of network propagation delay on forks and stale
blocks~\cite{nakamoto2008,decker2013information}. Subsequent analyses refined
statistical models of block arrivals and examined deviations from ideal Poisson
behavior under changing network conditions~\cite{bowden2018blockarrivals,
bowden2020modeling}. Separately, empirical studies of blockchain forks quantified
the frequency and duration of competing blocks and attributed prolonged forks to
network-level effects rather than to consensus failure~\cite{neudecker2019forks}.
These lines of work typically treat timing statistics, forks, and protocol
feedback as distinct phenomena, and often regard the exponential form as an
empirical regularity.

In contrast, we examine Bitcoin as a permissionless consensus system in which
time cannot be assumed as an external or synchronized input. We ask whether the
Poisson regime observed in block arrivals is merely incidental, or whether it is
actively sustained by protocol-level mechanisms under long-term variation in
hashrate and network conditions.

We analyze approximately 425{,}000 observed block arrivals from 2016 to 2024
\cite{blockarrivaltime}, spanning 211 complete difficulty epochs excluding 22
incomplete ones. Because Bitcoin does not enforce strict global clock
synchronization, block header timestamps are only loosely constrained and may
differ across miners. We therefore use arrival times observed at single nodes as
a consistent measurement basis. Using these data, we examine waiting-time
distributions, epoch-level rates, autocorrelations, and interval entropy.

Our analysis makes three empirical contributions. First, we demonstrate epoch-level exponential stability. Waiting times remain well-approximated by an exponential form across difficulty epochs, and estimated arrival rates stay near the protocol target over long horizons. Second, we show that discovery entropy concentrates in the high-entropy region of each interval. This indicates that blocks are typically found while uncertainty remains substantial. Third, we quantify distributed completion of entropy collapse using the survival function of fork durations. Most forks resolve rapidly, while a nonzero tail reflects propagation-bounded completion.

The paper is structured as follows. Section~II develops the theoretical foundations of Bitcoin’s temporal dynamics. Section~III provides empirical evidence for exponential arrivals and epoch-level stability. Section~IV analyzes interval-level entropy, its collapse at block discovery, and its implications for discrete temporal progression. Section~V concludes.

\section{Theoretical Framework for Bitcoin’s Discrete Temporal Dynamics}
\label{sec:theory}

This section formalizes a systems-level description of Bitcoin’s temporal dynamics using three components. First, proof-of-work is modeled as Bernoulli sampling in a vast hash space, which yields the Poisson arrival limit for block discovery. Second, interval-level uncertainty is quantified by entropy and is resolved at discovery, providing a discrete notion of temporal advancement at the level of observation. Third, difficulty adjustment is expressed as a feedback rule that regulates the rate of these discrete events across epochs.

Together, these elements define the theoretical structure that will be tested empirically in the following sections.

\subsection{Bernoulli Sampling, Difficulty, and the Poisson Arrival Limit}

Bitcoin’s proof-of-work mechanism operates through an immense number of
independent Bernoulli trials. Each hash attempt succeeds if its output lies
below the target threshold $T_{\mathrm{target}}$, which is determined by the
difficulty parameter
\begin{equation}
D = \frac{T_0}{T_{\mathrm{target}}}.
\end{equation}
where $T_0$ is a fixed reference target corresponding to difficulty one in the Bitcoin protocol. As difficulty increases, the admissible hash region contracts and the per-trial
success probability becomes
\begin{equation}
\theta = \frac{1}{D \cdot 2^{32}}.
\end{equation}

Let $H$ denote the global hashrate. By time $t$ the network performs
$N(t)=\lfloor Ht \rfloor$ independent trials, yielding the survival probability
\begin{equation}
\Pr[T > t] = (1-\theta)^{N(t)}.
\end{equation}
Bitcoin operates in the classical rare-event regime where $\theta \to 0$,
$N(t)\to\infty$, and $\theta N(t)$ remains finite. Taking this joint limit gives
\cite{ross1996stochastic}
\begin{equation}
\Pr[T > t] \to e^{-\lambda t},
\qquad
\lambda = H\theta \approx \frac{H}{D \cdot 2^{32}}.
\end{equation}

Thus $T$ is asymptotically exponential with rate $\lambda$, and block
arrivals follow a Poisson process. Since the per-trial success probability is
set by the difficulty target, this relation implies the hashrate estimator
\begin{equation}
H \approx \frac{1}{\theta\,\mathbb{E}[T]} \approx \frac{D\,2^{32}}{\mathbb{E}[T]}.
\end{equation}

The Poisson limit remains valid in the presence of multiple miners because their
search processes are effectively independent. Each miner constructs distinct
candidate block headers through transaction selection, ordering, coinbase data,
and nonce structure, thereby exploring a local possibility space $\Omega_m$.
Because each miner samples only a negligible fraction of the $2^{256}$ hash
domain, the overlap between any two such spaces is vanishingly small,
\begin{equation}
\Pr[\Omega_i \cap \Omega_j] \approx 0 \qquad (i \neq j).
\end{equation}
The network therefore samples from an effective global space
\begin{equation}
\Omega = \bigcup_{m=1}^M \Omega_m,
\end{equation}
and the time to block discovery is the minimum of $M$ independent exponential
waiting times. This minimum remains exponentially distributed with aggregate
rate $\lambda$, confirming that parallel mining preserves the Poisson arrival
structure. 

\subsection{Difficulty Feedback for Long-Horizon Stability}

Let $D_k$ denote the difficulty at the start of epoch $k$ and let
$T_{\mathrm{obs},k}$ be the observed duration of that epoch. The protocol aims
for an expected epoch duration of $T_{\mathrm{epoch}} = 2016 \times 600 \ \mathrm{s}$.
Because the Poisson limit yields an exponential waiting time with rate
$\lambda$, regulating the long-run arrival rate amounts to regulating $\lambda$
across epochs under changing hashrate.

Bitcoin updates difficulty via a discrete feedback rule based on the last
epoch's realized duration \cite{bitcoincore2025pow}
\begin{equation}
D_{k+1}
= D_k \frac{T_{\mathrm{epoch}}}{T_{\mathrm{obs},k}}.
\label{eq:difficulty-update}
\end{equation}

When blocks arrive faster than expected, $T_{\mathrm{obs},k} < T_{\mathrm{epoch}}$,
the update increases difficulty and reduces the per-trial success probability.
When blocks are slower, it decreases difficulty. This feedback stabilizes the
effective arrival rate across epochs and helps sustain the exponential regime
required for decentralized coordination.

\subsection{Entropy Collapse and the Granularity of Bitcoin Time}

For a fixed rate $\lambda$ within an interval, the probability that a block has
been discovered by time $t$ is
\begin{equation}
p(t) = 1 - e^{-\lambda t}.
\end{equation}

Let $\tau$ denote the discovery time within the interval. Under the longest-chain
rule, an honest node that observes a valid block immediately abandons its
ongoing search and begins the next interval from the newly adopted tip. The
interval-level uncertainty collapses locally and begins to collapse network-wide,
\begin{equation}
H(\tau^{-}) > 0,
\qquad
H(\tau^{+}) \rightarrow 0.
\end{equation}

Interval-level uncertainty can be quantified by the binary entropy
\begin{equation}
H(t)
= -p(t)\log_2 p(t)
  -[1-p(t)]\log_2(1-p(t)).
\label{eq:entropy}
\end{equation}
As an inter-arrival interval progresses, this entropy evolves deterministically (unimodally).
Under the longest-chain rule, an honest
node that observes a valid block immediately abandons its
ongoing search and begins the next interval from the newly
adopted tip. The interval-level uncertainty collapses locally and begins to collapse network-wide.

\begin{equation}
H(t^{-}) > 0,
\qquad
H(t^{+}) \rightarrow 0.
\end{equation}

This entropy collapse defines the generation of protocol-level time. Let
$0 < T_1 < T_2 < \dots$ denote successive block arrival times observed by an
honest node and define 
\begin{equation}
\Delta T_n = T_{n+1} - T_n.
\end{equation}
No new temporal information is produced by the protocol during the open
intervals $(T_n, T_{n+1})$. Bitcoin time therefore consists of the discrete
sequence of block arrival events $T = \{T_1, T_2, \dots\}$, rather than a
continuous timeline.

In this sense, the granularity of Bitcoin time is not imposed externally but
emerges endogenously from entropy accumulation and collapse. Each block
discovery both resolves uncertainty and advances the system’s internal temporal
state, establishing a non-continuous notion of time grounded in information
resolution rather than in physical clocks.

\section{Empirical Distribution of Inter-arrival Times}
\label{sec:interarrival}

\begin{figure}[t]
  \centering
  \includegraphics[width=\linewidth]{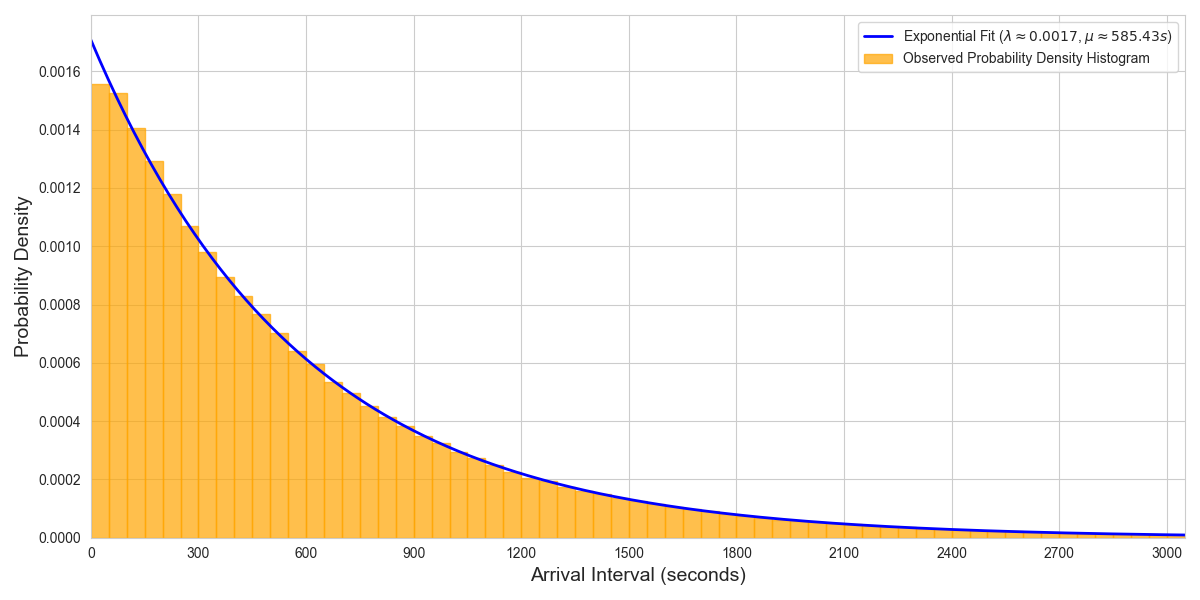}
  \caption{Histogram of Inter-arrival Intervals with Fitted Exponential density.}
  \label{fig:interval}
\end{figure}

This section examines whether the empirical sequence of inter-arrival intervals
conforms to the exponential model implied by proof of work. The analysis is based on observed block arrivals recorded from 2016 to 2024. 

\subsection{Inter-arrival Durations and Epoch Stability}

Let $\Delta T_1, \Delta T_2, \ldots, \Delta T_N$ denote the observed inter-arrival
intervals. The sample mean is
\begin{equation}
\bar{\Delta T}
  = \frac{1}{N} \sum_{i=1}^{N} \Delta T_i .
\label{eq:sample-mean}
\end{equation}

Figure~\ref{fig:interval} shows the empirical histogram of $\Delta T_i$ together
with the exponential density implied by the estimated rate
$\hat{\lambda} = 1 / \bar{\Delta T}$. The fitted distribution corresponds to a
mean inter-arrival interval of $585.43\,\mathrm{s}$. For visual clarity, the
figure truncates inter-arrival intervals above $3000\,\mathrm{s}$. Such extreme
intervals are rare, and their omission does not affect the shape of the central
distribution. The histogram exhibits the characteristic right-skewed form of an
exponential process. Consistent with the exponential model, the sample mean and
standard deviation are of comparable magnitude. These results indicate that the
global inter-arrival data conform closely to the exponential benchmark.

To examine stability across epochs, we compute an epoch-level mean inter-arrival
interval from block timestamp data for each 2016-block difficulty epoch. This
timestamp-based measure reflects the internal timing signal used by the
protocol’s difficulty adjustment mechanism.

\begin{figure}[t]
  \centering
  \includegraphics[width=\linewidth]{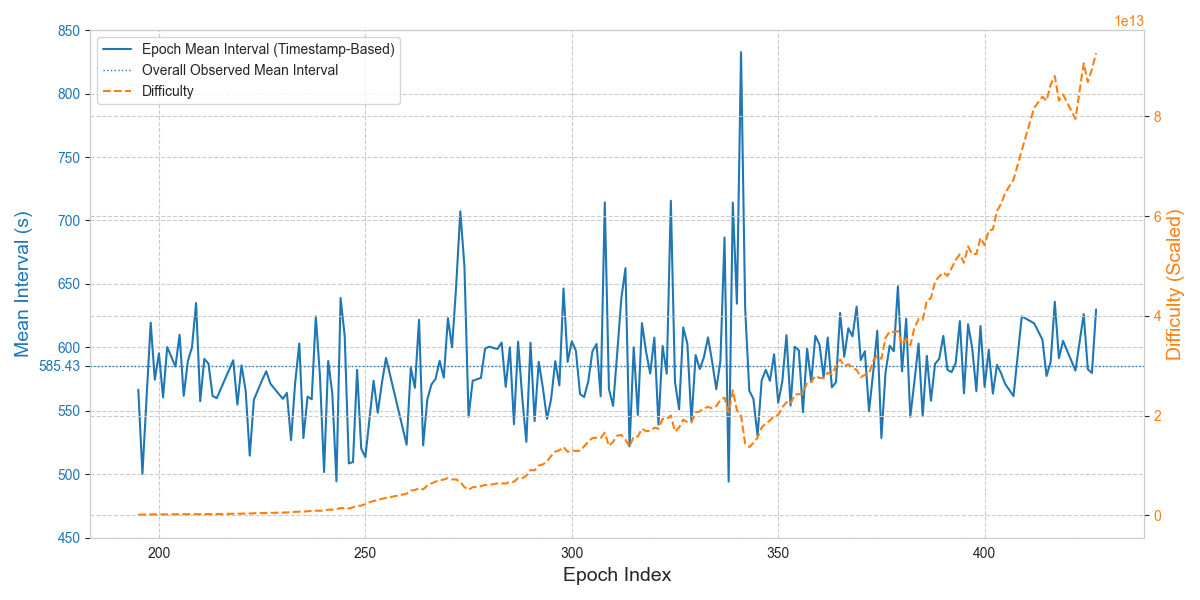}
  \caption{Epoch-mean Inter-arrival Intervals and Contemporaneous Difficulty.}
  \label{fig:epoch-lambda}
\end{figure}

Figure~\ref{fig:epoch-lambda} shows that epoch mean inter-arrival times cluster
tightly around the ten-minute target, with most values falling within
$600 \pm 100$ seconds and a mild tendency toward shorter intervals. The
horizontal reference line indicates the overall observed mean of 585.43
seconds.

This asymmetry reflects the interaction between discrete difficulty adjustment
and sustained hashrate growth. Within an epoch, difficulty is fixed, so increases
in network hashrate raise the effective arrival rate and bias inter-arrival
intervals toward shorter values later in the epoch. The subsequent retarget
reacts by increasing difficulty, but when hashrate continues to grow across
epochs, the feedback operates with delay on a moving target, allowing a
persistent deviation from the nominal ten-minute level.

To quantify this effect, we split each difficulty epoch into two halves of 1008
inter-arrival intervals and compute the mean duration in each half, denoted by
$\bar{T}_{\mathrm{early}}$ and $\bar{T}_{\mathrm{late}}$, respectively. Across
211 complete epochs, the late-half mean is systematically shorter than the
early-half mean. Averaged over all epochs, the early-half mean is
$\bar{T}_{\mathrm{early}} = 590.08\,\mathrm{s}$, while the late-half mean is
$\bar{T}_{\mathrm{late}} = 580.78\,\mathrm{s}$, yielding an average paired
difference of
$\bar{T}_{\mathrm{late}}-\bar{T}_{\mathrm{early}} = -9.30\,\mathrm{s}$ with a
95\% confidence interval $[-14.80\,\mathrm{s}, -3.81\,\mathrm{s}]$. The late
half is shorter in 132 out of 211 epochs (62.6\%), and the paired difference is
statistically significant (paired $t$-test $p=0.0010$).

Because difficulty is fixed within an epoch, the mean inter-arrival time is
inversely proportional to the effective hashrate. The observed bias therefore
implies a gradual within-epoch increase in hashrate, with an average ratio
\begin{equation}
\frac{H_{\mathrm{late}}}{H_{\mathrm{early}}}
\approx
\frac{\bar{T}_{\mathrm{early}}}{\bar{T}_{\mathrm{late}}}
\approx 1.016,
\end{equation}
corresponding to a $1$--$2\%$ increase from the early half to the late half. For comparison, the late-half mean of approximately $580\,\mathrm{s}$ is below
the $600\,\mathrm{s}$ target, indicating an effective arrival rate about
$600/580.78 \approx 1.033$ times faster than the protocol target on average.

Overall, these results highlight a separation of temporal scales in Bitcoin’s timekeeping mechanism. Difficulty is adjusted on a slower epoch scale through discrete feedback, constraining long-run timing behavior, while block arrivals fluctuate stochastically at the inter-arrival scale. Together, this behavior is consistent with a Poisson arrival process regulated by discrete adaptive feedback.

\subsection{Serial Independence}

If the exponential model holds, the sequence of inter-arrival intervals should
exhibit no serial dependence. Autocorrelations are computed using
\begin{equation}
\rho(k)
  = \frac{\sum_{i=1}^{N-k}
      (\Delta T_i - \bar{\Delta T})
      (\Delta T_{i+k} - \bar{\Delta T})}
    {\sum_{i=1}^{N}
      (\Delta T_i - \bar{\Delta T})^2}
\label{eq:acf}
\end{equation}
for lags $k = 1, 2, \ldots, 20$.

\begin{figure}[t]
  \centering
  \includegraphics[width=\linewidth]{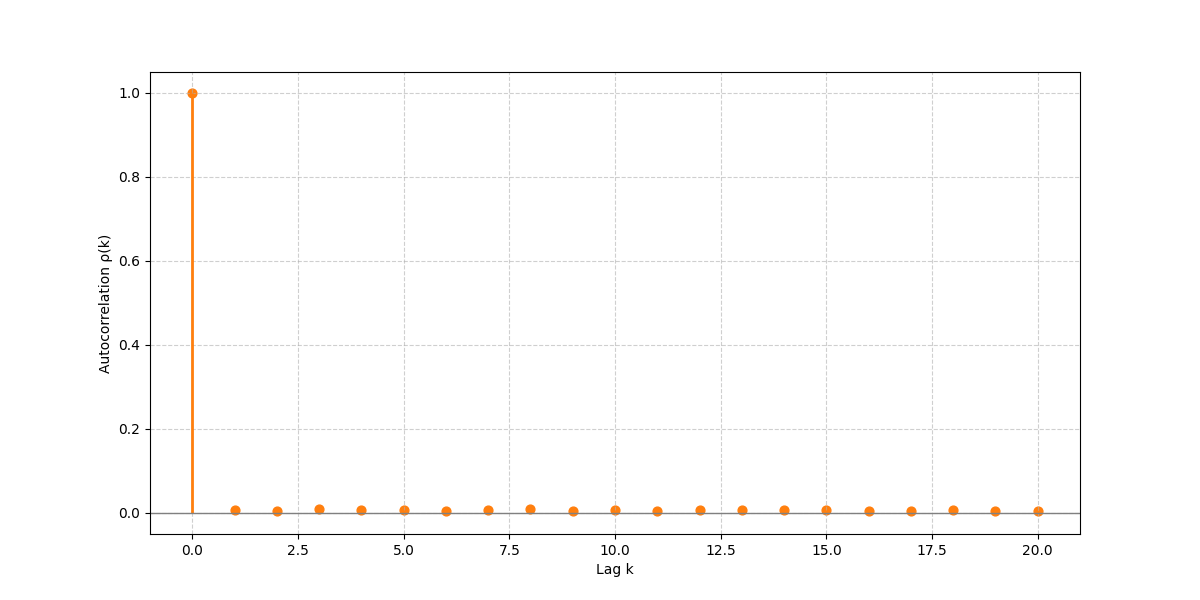}
  \caption{Autocorrelation of Inter-arrival Intervals for Twenty Lags.}
  \label{fig:acf}
\end{figure}

Figure~\ref{fig:acf} reports the sample autocorrelation function for the trimmed
inter-arrival intervals. To reduce the influence of extreme outliers, we remove
the top and bottom one percent of durations. Autocorrelations at lags 1 through
20 fluctuate tightly around zero and remain well within sampling bounds, with no
detectable serial dependence. This behavior is consistent with independent
exponential waiting times and a Poisson arrival process driven by independent
hash attempts.

These results characterize the external timing of block arrivals observed by a
participant, but they do not describe how temporal uncertainty accumulates
within an interval or how it resolves at discovery. To capture these internal
dynamics, Section~\ref{sec:entropy} introduces an interval-level entropy measure
and shows how collapse events define the discrete temporal progression of the
system.

\section{Interval-level Entropy and Collapse}
\label{sec:entropy}

Entropy provides an internal representation of uncertainty within each block
interval. While the arrival probability $p(t)$ increases monotonically, the
corresponding binary entropy $H(t)$ follows a concave profile, peaking at
$p(t)=1/2$ and remaining elevated over much of the interval
\cite{coverthomas}. For Bitcoin’s target interval of approximately 600 seconds,
this high-entropy regime is reached well before the expected block arrival,
indicating that substantial uncertainty persists throughout most of the
interval.

As formalized in
\cite{chen2025noncontinuous}, collapse completion occurs only upon block discovery. Under the
longest-chain rule, an honest node that observes a valid block immediately
abandons its ongoing search and begins the next interval from the newly adopted
tip. This discrete reduction of uncertainty constitutes entropy collapse and
defines the advancement of Bitcoin’s non-continuous time.

\begin{figure}[t]
\centering
\includegraphics[width=\linewidth]{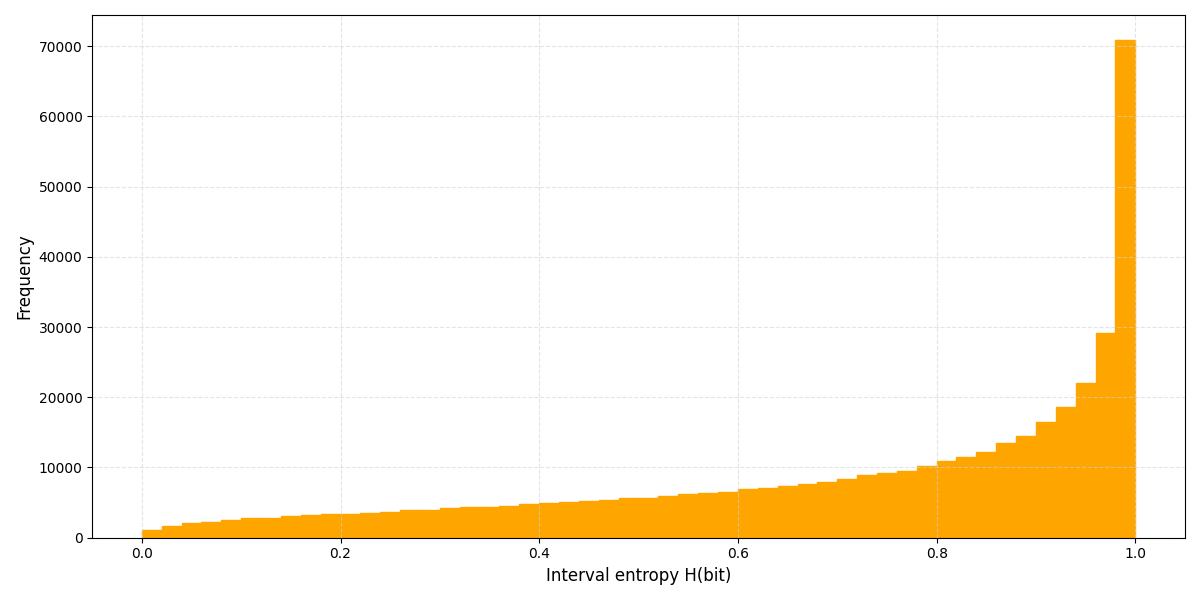}
\caption{Histogram of Entropy at Block Discovery}
\label{fig:entropy-hist}
\end{figure}

Building on this framework, we evaluate the entropy of each observed block
interval at its realized duration using a global mean arrival rate estimated
from the full sample. Figure~\ref{fig:entropy-hist} shows the resulting entropy
distribution. The histogram is sharply concentrated near the maximal entropy
region, indicating that most block intervals terminate while uncertainty
remains high. Only a small fraction of unusually short or long intervals appear
in the low-entropy tails.

This empirical pattern provides direct observational support for the
entropy-collapse mechanism, showing that block intervals typically terminate
while uncertainty remains high and that entropy collapse is initiated by the
discrete occurrence of a valid proof-of-work solution rather than by gradual
probabilistic resolution. From an informational perspective, each block
discovery terminates one interval and initiates the collapse of the uncertainty
accumulated since the previous block. The evolution of this uncertainty within
an interval is governed by the block arrival rate, which determines how rapidly
entropy accumulates over time and is the statistical quantity estimated and
regulated by the difficulty-adjustment mechanism.

Crucially, in a distributed network this collapse is not completed
instantaneously. While entropy collapse is initiated locally at block
discovery, residual uncertainty may persist across the network due to
propagation delays and competing discoveries. The completion of entropy
collapse therefore unfolds over a finite, propagation-bounded interval, giving
rise to temporary forks and distributed resolution.

\subsection{Distributed Completion of Entropy Collapse}

In a distributed network, entropy collapse unfolds as a process rather than as
an instantaneous event. Because nodes do not observe block discoveries
simultaneously, collapse proceeds through a propagation-bounded completion
phase during which multiple candidate blocks may temporarily coexist. When
multiple valid blocks are discovered at the same height within this window,
each represents a local realization of collapse, while global uncertainty
persists until one branch dominates through cumulative work and the
longest-chain rule \cite{nakamoto2008}. Temporary forks thus provide an
observable manifestation of the distributed completion of entropy collapse,
rather than a breakdown of consensus.

To quantify this completion phase, we define the fork duration as the
time interval between the first observed valid block at a given height and the
last competing block observed before convergence to a single branch. Fork
duration measures how long residual uncertainty persists after collapse is
initiated. It is distinct from block inter-arrival time, which characterizes
the spacing of successive accepted blocks on the canonical chain. Fork duration
thus provides an observable proxy and a lower bound on collapse completion time,
as latent propagation delays or unseen orphan blocks may extend the true
collapse completion process beyond what is directly measurable.

Empirically, fork durations are analyzed using the survival function
\begin{equation}
\hat{S}(\tau)
= \Pr(\mathcal{D} \ge \tau)
= \frac{1}{N} \sum_{i=1}^{N} \mathbf{1}\!\left(\mathcal{D}_i \ge \tau\right),
\label{eq:ccdf}
\end{equation}
where $\mathcal{D}$ denotes the fork-duration random variable,
$\{\mathcal{D}_i\}_{i=1}^{N}$ are observed realizations of $\mathcal{D}$, and
$\tau \ge 0$ denotes a duration threshold measured as elapsed time since the
initiation of a fork.

\begin{figure}[t]
  \centering
  \includegraphics[width=0.9\linewidth]{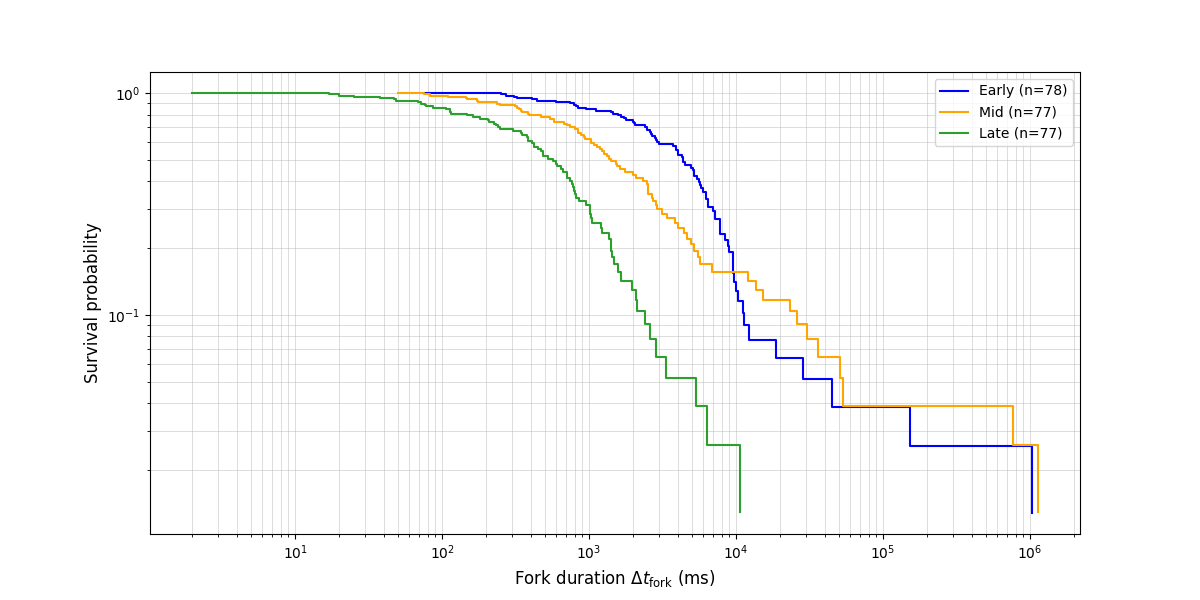}
  \caption{Empirical Survival Function of Fork Durations Following Block Discovery.}
  \label{fig:fork_duration_ccdf}
\end{figure}

Using the publicly available fork dataset collected via direct network
monitoring~\cite{neudecker2019forks}, Figure~\ref{fig:fork_duration_ccdf} plots
the empirical survival function of fork durations on a log--log scale. Consistent
with prior measurements, fork events are rare and form a sparse sequence ordered
by blockchain height rather than a temporally contiguous time series, with
non-fork intervals omitted between successive observations~\cite{decker2013information}.

To isolate structural changes under evolving network conditions, we partition
the data by blockchain height into three consecutive segments containing 78, 77,
and 77 fork intervals, respectively. This stratification reveals a systematic
compression of fork durations over time. All segments exhibit a steep decay,
indicating that most forks resolve rapidly, while the presence of a nonzero tail
confirms that resolution is not instantaneous but unfolds over a finite,
propagation-bounded interval. The late segment shows a markedly faster decay and
a substantially reduced tail relative to earlier segments, consistent with
improved network propagation and a reduced window for competing discoveries.

While the dependence of fork duration on network conditions has been documented
previously~\cite{neudecker2019forks}, our contribution is to reinterpret these
observations within an entropy-collapse framework. In this view, proof-of-work
competition determines when collapse is initiated, whereas network propagation
governs how rapidly collapse is completed in a distributed setting. Bitcoin
thereby constructs a coherent and verifiable temporal order without requiring
instantaneous global agreement or external synchronization.

\subsection{Implications for Temporal Stability}

When the ratio $H/D$ remains approximately constant, the arrival rate implied by
the exponential limit remains stable. Under this condition, the system
maintains a consistent temporal scale without reference to any external clock.
The combination of entropy collapse at the interval scale and mean reversion at
the epoch scale implies that a proof-of-work system with stationary fundamentals
can function as an autonomous time source. It produces a verifiable sequence of
discrete uncertainty-resolving events that can be interpreted consistently by
all participants.

This property is particularly relevant in open and distributed environments
where no trusted or synchronized external clock is available. In such settings,
time cannot be assumed as a shared input but must be constructed from observable
events. An arrival process defined by entropy collapse provides such a mechanism
by establishing a common temporal reference through ordered and verifiable
events, rather than through dependence on external time authorities.
This provides a consistent explanation of Bitcoin’s non-continuous time, both
at the level of block discovery and under distributed observation.

\section{Conclusion}

Prior to Bitcoin, Bernoulli and binomial processes were well understood in
theory but had not been realized as an open and independently verifiable system
operating in the real world. Bitcoin instantiates such a system through
probabilistic sampling, cryptographic verification, and adaptive difficulty
control. Empirically, global hash trials operate in the Bernoulli limit and
produce exponential inter-arrival intervals consistent with Poisson theory.
Uncertainty evolves within each interval and is resolved at block discovery,
while difficulty adjustment stabilizes the event rate across epochs without
suppressing short term randomness. In a distributed network, however, entropy
collapse is not completed instantaneously. Empirical observations of temporary
forks indicate that residual uncertainty may persist for a short
propagation-bounded interval following block discovery before a single history
becomes dominant. Together, these results show that Bitcoin maintains a stable yet non-continuous
temporal structure without reliance on any unified external clock. More
importantly, they illustrate how a permissionless distributed system, in which
no trusted reference time can be assumed, can construct a verifiable temporal
order endogenously from probabilistic events and protocol-level information.

\section*{Acknowledgments}

The author thanks the Bitcoin developer and research communities for making block and difficulty data publicly accessible. Any errors remain the author's own.

\bibliographystyle{IEEEtran}

\begin{thebibliography}{99}

\bibitem{chen2025noncontinuous}
B. Chen.
Bitcoin: A Non-Continuous Time System.
arXiv preprint arXiv:2501.11091, 2025.

\bibitem{nakamoto2008}
S. Nakamoto.
Bitcoin: A Peer-to-Peer Electronic Cash System.
2008.
Available: \url{https://bitcoin.org/bitcoin.pdf}

\bibitem{decker2013information}
C. Decker and R. Wattenhofer.
Information propagation in the Bitcoin network.
In \textit{IEEE P2P 2013 Proceedings}, pages 1--10, 2013.

\bibitem{bowden2018blockarrivals}
R. Bowden, S. Keil, A. Gervais, S. Matetic, and S. Capkun.
Block arrivals in the Bitcoin blockchain.
arXiv preprint arXiv:1801.07447, 2018.

\bibitem{bowden2020modeling}
R. Bowden, H. P. Keeler, A. E. Krzesinski, and P. G. Taylor.
Modeling and analysis of block arrival times in the Bitcoin blockchain.
\textit{Stochastic Models}, 36(4):602--637, 2020.

\bibitem{neudecker2019forks}
T.~Neudecker and H.~Hartenstein,
``An Empirical Analysis of Blockchain Forks in Bitcoin,''
in \emph{Financial Cryptography and Data Security (FC 2019)}, pp.~84--92, 2019.

\bibitem{blockarrivaltime}
Bitcoin-data.
Bitcoin Block Arrival Time Dataset.
Accessed 2025.
Available: \url{https://github.com/bitcoin-data/block-arrival-times}

\bibitem{ross1996stochastic}
S. M. Ross,
\textit{Stochastic Processes}.
John Wiley and Sons,
1996.

\bibitem{bitcoincore2025pow}
Bitcoin Core Developers.
Bitcoin Core Source Code: Difficulty Adjustment Logic (\texttt{src/pow.cpp}).
GitHub Repository, 2025.
Available: \url{https://github.com/bitcoin/bitcoin}

\bibitem{coverthomas}
T.~M. Cover and J.~A. Thomas,
\emph{Elements of Information Theory},
John Wiley \& Sons, 1999.

\end{thebibliography}

\end{document}